\documentclass[preprint,3p,times,twocolumn]{elsarticle}

\usepackage{amssymb}
\usepackage{booktabs}
\usepackage{enumitem}

\usepackage{multicol}
\usepackage{multirow}

\usepackage{wasysym}
\usepackage{booktabs}
\usepackage{float}
\usepackage{url}

\usepackage{bibentry}
\usepackage{supertabular}
\usepackage{longtable}

\usepackage{array}                  
\newcolumntype{L}[1]{>{\raggedright\let\newline\\\arraybackslash\hspace{0pt}}m{#1}}
\newcolumntype{C}[1]{>{\centering\let\newline\\\arraybackslash\hspace{0pt}}m{#1}}
\newcolumntype{R}[1]{>{\raggedleft\let\newline\\\arraybackslash\hspace{0pt}}m{#1}}
\newcolumntype{J}[1]{>{\let\newline\\\arraybackslash\hspace{0pt}}m{#1}}

\setlist{nosep}

\journal{  }

\begin{document}
\begin{frontmatter}



\title{On the Usage of Psychophysiological Data in Software Engineering: \\An Extended Systematic Mapping Study}


\author[add1]{Roger Vieira}
\ead{rogervi@edu.unisinos.br}
\author[add1]{Kleinner Farias \corref{corr1}}
\ead{kleinnerfarias@unisinos.br }

\address[add1]{Graduate Program in Applied Computing (PPGCA), \\ University of Vale do Rio dos Sinos (Unisinos), \\São Leopoldo, Rio Grande do Sul, Brazil}

\cortext[corr1]{Corresponding author}

\begin{abstract}
In recent years, many studies have applied wearable devices to capture psychophysiological data from software developers. However, the current literature lacks investigations that classify the studies and point out gaps to be explored. This article, therefore, seeks to present a comprehensive overview of the literature by classifying and creating a systematic map of the works. Besides, it seeks to pinpoint research gaps, challenges, and trends. Based on widely known guidelines, a systematic mapping of the literature was designed and run to answer eight research questions. After applying a careful filtering process, we selected 27 representative studies from a sample of 2,084 potentially relevant works retrieved from seven digital libraries. The main results are: a classification scheme of the published studies was produced; there is no predominance of the devices used to capture psychophysiological data; over 50\% of the studies have explored indicators related to mental states and neural activity; and 80\% have analyzed composite data to understand the cognitive load and in the context of understanding debugging programs and strategies. Our findings can benefit researchers and students by creating a systematic map of the literature, being a starting point for future research.
\end{abstract}




\end{frontmatter}


\section{Introduction}

Software development is an intellectual process, where every developer applies knowledge grounded in a set of logical premises to produce solutions designed to real-world problems. The development needs professionals to be focused and deeply engaged in their work, paying attention to every detail which offers a threat to impact negatively the produced solution~\cite{Duraes2016}. Moreover, the development of real-world applications has happened in increasingly unstable business environments~\cite{oliveira2018brcode}, requiring collaborative work~\cite{lucas2017collabrdl,de2019umlcollab,neto2019supporting,mclyndon2019umlcollab}, development effort from different developers\rq perspectives, including the effort to integrate software artifacts~\cite{farias2010empirical,farias2015evaluating,farias2013analyzing,farias2014effects,farias2012evaluating}, maintain source code~\cite{d2020effects}, and evolve software architecture~\cite{farias2010assessing}.


Nowadays, studies aim to understand which cognitive processes are related to software development tasks, using technologies as Electroencephalography (EEG) and Functional Magnetic Resonance Imaging (fMRI) \cite{Siegmund2014, Duraes2016, Floyd2017,manica2018effects,vieira2020cognide}. In these studies, the authors advocate that it is possible to find a relation between defect production in software artifacts and the developer\rq s cognitive processes presented when source code was written, aiming the development of mechanisms capable of priory low confidence code refactoring and even prediction of defective software artifacts. Manica et al.~\cite{manica2018effects} investigate how software developer\rq s effort and affective states can be affected by model composition techniques. Vieira and Farias~\cite{vieira2020cognide} introduce a psychophysiological data integrator approach for for the Microsoft Visual Studio code.

The advent of microelectronics brought smaller, faster, and low energy profile processors, enabling the usage of mobile and accessible devices to collect biological signals, including in several academic experiments. This is the case of Mindwave Mobile 2 \cite{Rostami2015} and Emotiv EPOC and EPOC+ \cite{Radevski2015}, devices that are used to increase comprehension upon cognitive processes involved in the coding activities. Allied to the extensibility of current Integrated Development Environments, the usage of these devices might enable the development of new functionalities enabling the identification of developer\rq s behavior patterns about their productivity and produced software artifacts quality \cite{Fritz2016, Radevski2015, Zayour2013}.

In the last decade, psychophysiological data has been widely used as an improvement resource in the software engineering context. However, the current literature lacks a formal classification and a systematic mapping of articles already made and published approaching this topic \cite{Floyd2017, Gonc2019}. Consequently, a detailed understanding of the usage of that data remains limited. Grounded on such an argument, the present study aims to answer the following research question: ``How are the psychophysiological data are being used as improvement resources in the software engineering context?''

Therefore, this work seeks to classify and provides a thematic analysis of studies that have already been published upon the usage of psychophysiological data as an improvement resource in the software engineering context. To fulfill that, a systematic mapping was designed and performed, following a set of well-established assumptions aiming to answer eight research questions. An initial number of 2,084 studies were collected within 7 digital databases, of which 27 were selected to answer the previously formulated research questions. As result, it notices that most studies are outlined as a controlled experiment, characterized by the usage of devices to collect psychophysiological data and integrated development environments (IDEs), intended to analyze developer\rq s psychophysiological attributes, mainly when it was applied in the comprehension of source code and debugging strategies. Besides, it was not identified mechanisms capable of enabling support for data integration and analysis natively within IDEs. Lastly, this study has identified some challenges, which would be explored by the scientific community for the next years.


This work is an extension of an initial literature review~\cite{ROGER2020SBSI}. This extended article differs by (1) exploring eight research questions, as opposed to 6 questions, (2) presenting a description of essential concepts related to psychophysiological data and integrated development environments, (3) detailing the methods used for the extraction of categories and their correct application for the classification of the answers obtained during the filtering stage, and (4) introducing the results in more detail. The presentation of the theoretical framework used, coupled with the addition of research questions and a more in-depth analysis of the results, therefore culminated in a better presentation of the identified findings and challenges, also enabling the making of a heat map capable of illustrating possible gaps to be treated as opportunities in the development of future studies.

Initially, this study presents a literature revision upon theoretical background adopted in its development (Section \ref{sec:background}), also enumerates and synthesizes the related works identified about the topic (Section \ref{sec:related_works}). Next, the adopted steps as the methodology for systematic mapping development are detailed (Section \ref{sec:methodology}). After that, the performing of selection, filtering and selected studies data extraction are showing up (Section \ref{sec:filtering}), like the found results (Section \ref{sec:results}) and the identified challenges are discussed grounded by the current literature (Section \ref{sec:discussion}). Finally, conclusions, final thoughts, and future works are conferred in Section \ref{sec:conclusion}.

\section{Background}
\label{sec:background}


This section presents the main concepts needed to understand the literature mapping study. The concepts are derived from articles already published (discussed in Section~\ref{sec:related_works}) on the topic of literature review considering the use of psychophysiological data. 

\subsection{Psychophysiological Metrics and Data}


The psychophysiological data\footnote{In academic literature, we often find the term \textit{biometric data}, however, in this article, the authors have adopted the term \textit{psychophysiological data} to avoid ambiguity across the process to assign an identity to a subject based on the biological data (aka.: \textit{biometry})} are measurements produced from metrics concerned with quantifying psychophysiological aspects of the human being. These measurements seek to quantify human responses to external stimuli in specific situations, for example, software developer\rq s cognitive load produced when reading the source code in a situation of pressure to deliver a software product to the market. Some studies~\cite{li2015, Andreassi2007} have explored the variation of measurements to better understand the mental states generated when human beings are exposed to situations produced through controlled experiments. The literature has explored the use of various metrics in different contexts of software development~\cite{Fritz2014,Fritz2016,oliveira2008quantitative,farias2011evaluating,de2012empirical,guimaraes2010analyzing}.

\subsubsection{Eye Metrics}


The eye metrics are often obtained using devices known as \textit{eye-trackers}. These devices use laser beams or infrared light to measure a set of attributes related to the subject\rq s eyeball, such as eye movement trajectory, pupil dilation, focus, eye gaze, and eye blinking frequency.


When applied in specific contexts, the collected data can be used to indicate psychophysiological processes, such as cognitive load, excitation, valency, attention, anxiety, and stress. The literature points to the viability of the experimental use of these devices in various contexts, such as monitoring the execution of software development tasks~\cite{Fritz2014}, performing software debugging strategy \cite{Lin2016}, and the difficulty of learning a new language~\cite{Hejmady2012}.

\subsubsection{Brain Metrics}

The brains' neural activity while executing several tasks chains a series of fluctuations in potential difference (or voltage), all long over the cranial surface, which can be measured using a technique called \textit{electroencephalography} (EEG) \cite{Fritz2016, Andreassi2007}. 

Researchers have identified the possibility to translate the neural activity and its voltage fluctuations in a wave spectrum grouped by different frequency bands, being able to be related to a respective set of activities (Table \ref{tab:waves}). Alpha wave ($\alpha$), for instance, often can be observed in subjects who are in states of meditation or relaxation, being attenuating as physical or mental activity intensity increases \cite{Andreassi2007, Fritz2016}. Electrodermal activity, as well as skin's surface temperature measurement, has already been correlated upon psychological processes, like excitation and some specific emotions \cite{Fritz2016}.

\begin{table}[t]
    \small\
    \caption{Brainwaves spectrum and its frequently related activities.}

    \label{tab:waves}
    \centering
    \begin{tabular}{ p{0.1\linewidth} p{0.2\linewidth} p{0.2\linewidth} p{0.3\linewidth}}
        \toprule
        \footnotesize
         \textbf{Wave} & \textbf{Symbol} & \textbf{Band} & \textbf{Relationship}\\ 
        \midrule
        Alpha       & $\alpha$  & 8 -- 12Hz    & relaxation, meditation, reflection\\ 
        Beta        & $\beta$   & 12 -- 30Hz   & alert state, focus\\ 
        Gamma       & $\gamma$  & 31 -- 120Hz  & high brain activity, learning\\ 
        Delta       & $\delta$  & 0,4 -- 4Hz   & deep sleep\\ 
        Theta       & $\theta$  & 4 -- 7Hz     & drowsiness, meditation\\ 
        \bottomrule
    \end{tabular}
\end{table}

As well as voltage fluctuations on the cranial surface, neural activities trigger biological processes according to the execution of both physical and mental activities, such processes as the brain blood and oxygen supply, can be measured employing techniques like \textit{functional Near-Infrared Spectroscopy} (fNIRS) and \textit{functional Magnetic Resonance Imaging} (fMRI). Instead of EEG, fNIRS and FMRI are capable of generating clear images of brain activity in well-defined regions, enabling its relation to cognitive and biological processes \cite{Fishburn2014, liLu2009, Solovey2015DesigningII}.

\subsubsection{Skin Metrics}

The set of metrics related to a variation on skin's surface properties, such as temperature and \textit{Electrodermal Activity} (EDA) is known by \textit{Skin metrics}, the last one is often found in the literature under the term \textit{Galvanic Skin Response} (GSR). The EDA sensor devices usually collect data from the skin\rq s electric resistance as a response to some external stimulus, for instance, when a subject is in a state of excitation, sweat glands rise its activity, triggering an increase of sweat production, consequently, reducing the electrical resistance of the skin, enabling a better electrical current conductibility across the skin \cite{Fritz2016, haag2004}.

\subsubsection{Cardio-respiratory Metrics}

In the revision process of the studies used as theoretical grounding, there were identified four main metrics associated to cardiorespiratory aspects presented by the subjects analyzed in the experiments: \textit{Blood Volume Pulse} (BVP), \textit{Heart Rate} (HR), \textit{Heart Rate Variability} (HRV) and the \textit{Respiratory Rate} (RR). The heart rate refers to the number of cardiac cycles (systolic and diastolic movements) performed by cardiac muscle on the period of one minute, on the other hand, the heart rate variability points to time variation between two consecutive cardiac cycles. As just the HR, the respiratory rate stands for the cycles performed by the lungs to provide oxygen to the body. The blood volume pulse, also found in literature as just ``pulse'', measures the blood flow in human body specific areas according to changes drove by \textit{Sympathetic Nervous System} (SNS), especially when a subject is exposed to stressful situations \cite{Andreassi2007, Fritz2016}. The presented metrics are strongly linked to cognitive load and several emotions \cite{Fritz2016}.

\subsection{Cognitive Load Theory}

The human body, besides constantly provides physical resources, does also with the called mental resources -- as memory, attention, and reasoning -- aiming the maintenance of the subject's vital functions. When necessary, such resources can be also allocated in specific ways, to enable the execution of harder tasks started by the subject, for instance, the resolution of logical problems faced by him in the software development process.

To the number of mental resources, especially the working memory, which appears while a mental effort is necessary to perform a set of cognitive activities - -both simple and complex ones -- it is attributed the term \textit{Cognitive Load} \cite{Sweller2011CognitiveTheory}. Due to its kind, the Cognitive Load can still be split into three specialized categories:

\begin{enumerate}
    
    \item \textbf{Intrinsic Cognitive Load}: it is related to the level of difficulty associated with a simple instructional topic. All instructions have an inherent difficulty level associated with it, which cannot be modified by the instructional agent. However, the mental schema used to represent a task might be fractioned in smaller pieces, enabling the possibility to be handled singly and posteriorly regrouped without compromising semantically the original schema.
    
    \item \textbf{Extraneous Cognitive Load}: such kind of cognitive load can be generated according to the way an informational content is presented by an agent to one who will perform an activity. Different approaches to present information can generate distinct levels of extraneous cognitive load.

    \item \textbf{Germane Cognitive Load}: different from such other cognitive load categories, the germane cognitive load can be considered a type of \textit{artificial} cognitive load, once it results from the process of mental schema creation used to represent a collection of instructions, such as automation, analogies or event deconstruction and construction of concepts, intending to provide easiness in the comprehension and learning regarding a specific topic.
    
\end{enumerate}

In short, the developer's cognitive load, for instance, can be affected by many factors, such as the source code presentation within integrated development environments, his expertise regarding a specific programming language, his knowledge about-faced errors, and his solution comprehension as a whole \cite{Fritz2016}.

\subsection{Integrated Development Environment}

The Integrated Development Environments (IDEs) can be considered as the main set of tools used by developers, for both creation tasks and source code maintenance. This set of tools holds source code editors, file navigation, compilers, and debuggers, artifacts refactoring assistants, and tools for syntax highlighting, such as other functionalities \cite {Snipes2015}. From a developer perspective, besides the expertise across the programming language which are being used and the full knowledge regarding logical aspects involved in the problem, it is essential his domain over the tools used in the development process \cite {Astromskis2017}.

In the last years, some authors have conducted researches about the usage of IDEs by developers, both in the software development process \cite{Astromskis2017, Zayour2013}, and its efficacy as a debugging tool \cite{Zayour2013}. Moreover, domain-specific modeling environments have been proposed to allow developers to work with semantically richer IDEs~\citet{farias2009mas, gonccalves2015mas,gonccalves2011mas,gonccalves2010modelagem}. Usually employing eye-trackers to collect quantitative data, the authors evaluate aspects like usage easiness, user graphical interface organization and disposition, the set of available tools, such other resources according to the solution provider. Looking for the evolution of such tools and to facilitate the development of related studies, some author has been produced articles aiming to perform a better way to integrate biometric data --- such as these collected by eye-trackers -- within integrated development environments, in a way that data can be analyzed and related to the generated source code, or even, to the developer's interaction while executing his tasks, with no need of external tools \cite{Guarnera2018}.

\subsection{Psychophysiological Data in Software Engineering}
Psychophysiological data has great applicability in understanding what experiences developers are subjected to while performing their tasks, during an intense, fully development-focused work cycle, as well as when facing some sort of difficulty understanding a piece of code or the process. solution design logic \cite{Fritz2016}.

In the context of Software Engineering, the use of psychophysiological data has been emerging as it refers to the understanding of cognitive processes involved in software development and their correlates \cite{Fritz2016}. Studies in this area have explored from devising new low-cost sensor devices that are capable of reliably measuring psychophysiological signals to examining how novice and experienced developers understand source code writing \cite{Lee2016, Szu2013} and how this can help improve software artifact production \cite{Radevski2015}.

Despite the increase in the production of studies in this area, some authors advocate the low consistency and lack of consensus presented in them, thus seeking to conduct surveys that bring a current and consolidated view around the use of psychophysiological data in the context of Software Engineering \cite{Gonc2019, Obaidellah2018}. Through the development of mapping and systematic reviews, the authors aim to identify gaps in the theme that may serve as input for the preparation of new studies able to address the problems presented.

\section{Related Work}
\label{sec:related_works}

This section presents a comparative analysis of this article with the current literature. To this end, several articles were identified after applying the search string ``Psychophysiological Data AND Software Engineering '' to Google Scholar \footnote{https://scholar.google.com/}, seven of which were selected for similarity to the theme explored in this study. Such studies are briefly analyzed below.

\subsection{Synthetic Analysis}

Intending to provide a wide point of view involving all related works, each of them was read and a synthetic analysis was performed to present a brief resume of them. The following is a list of a synthetic analysis of each related work.

\begin{itemize}
	\item\textbf{ \citet{Fritz2014}}: it presents a new approach to the classification of difficulty in understanding source code that was investigated using biometric data associated with psychophysiological states. Through an experiment involving 15 developers, it was used \textit {Eye-trackers}, electrodermal activity sensors (EDA), and electroencephalograph (EEG) to verify how the collected data could provide a form of prediction when a developer would consider a task as being difficult.
	
	\item \textbf{\citet{Fritz2016}}: the dissertation was based on a series of controlled experiments carried out in previous studies, which applied different sensors for data collection. They also discussed the strategies adopted for data collection, preparation, transformation, loading, and analysis, as well as to elucidate the machine learning algorithms applied to extract relevant features to the objectives of each study.
	
	\item \textbf{\citet{Fucci2019}}: the authors replicated experiments previously presented in the literature \cite{Floyd2017}, using low-cost sensor devices. To do so, they performed measurements using EEG, EDA, and BVP  on a sample of 28 Computer Science students while performing tasks involving the comprehension of texts written in natural language and source code.
	
	\item \textbf{\citet{Gonc2019}}: a systematic mapping was developed to identify gaps in previous works that addressed the study of cognitive load in the context of Software Engineering. Thirty-three related articles that intersected the themes of cognitive load, software engineering, and use of sensor devices for psychophysiological data collection were analyzed.
	
	\item \textbf{\citet{Gui2019}}: a literary survey was conducted between 2007 and 2017 on the state of the art about the use of brain biometric data, the types of sensors used, how these data were collected, their application contexts, and statistical approaches, and machine learning to extract features around the use of such data.
	
	\item \textbf{\citet{Radevski2015}}: a conceptual architecture model was proposed aiming at real-time data acquisition from low-cost EEGs. A pilot test was also performed using Emotiv EPOC and EPOC + devices to evaluate features such as usability and reliability during application in experiments. Moreover, the authors have addressed the ethical implications involved in collecting real-time psychophysiological data for developer productivity assessment.
	
	\item \textbf{\citet{Rostami2015}}: through an experimental study, the cognitive load and familiarity of a group of Computer Science students were measured regarding their interaction with Integrated Development Environments, using a low-cost EEG device to collect data. psychophysiological data. The authors also pointed out the advantages of EEG over similar techniques and the technical limitations of low-cost sensor devices.
\end{itemize}

\subsection{Comparative Analysis and Opportunities}

Five Comparison Criteria (CC) were defined to assist in the process of identifying similarities and differences between the proposed work and the selected articles. This comparison is crucial in making the process of identifying research opportunities using objective rather than subjective criteria. The criteria are described below:

\begin {itemize}
        \item \textbf{Systematic Mapping (CC01):} studies that developed a systematic mapping of the literature aiming at acquiring a vision of the current state of the art.
        \item \textbf{Data Integration (CC02):} papers that researched how psychophysiological data are being supported by integrated development environments.
        \item \textbf{Psychophysiological Metrics (CC03):} these categories are studies that researched biological metrics that were linked to cognitive processes.
        \item \textbf{Applications in Software Engineering (CC04):} studies that sought to evaluate how psychophysiological data are applied in Software Engineering.
        \item \textbf{Integrated Development Environments (CC05):} researches related to integrated development environments, both around its use and its improvement.
\end{itemize}

Table \ref{tab:comp} presents the comparison of the selected works, contrasting them with the proposed work. Some research gaps and opportunities are noted:

\begin{enumerate}
    \item The proposed work was the only one that completely met all the comparison criteria;
    \item Psychophysiological metrics and applications in Software Engineering were the most explored themes, while systematic mapping and integrated development environments were the least explored;
    \item No studies have explored in detail the use of psychophysiological data in the context of integrated development environments.
\end{enumerate}

\begin{table}[ht]
    \footnotesize
    \caption{Comparative analysis among related works}
    \label{tab:comp}
    \centering
    \begin{tabular}
    {
        J{0.35\linewidth}
        C{0.05\linewidth}
        C{0.05\linewidth}
        C{0.05\linewidth}
        C{0.05\linewidth}
        C{0.05\linewidth}
    }
        \toprule
			\multicolumn{1}{c}{\multirow{2}[2]{*}{\textbf{Related Work}}} & \multicolumn{5}{c}{\textbf{Comparison Criteria}}
			\\\cline{2-6}
			& \textbf{CC1} & \textbf{CC2} & \textbf{CC3} & \textbf{CC4} & \textbf{CC5}  \ \\

        \midrule
        Proposed Work 	& $\CIRCLE$ & $\CIRCLE$ & $\CIRCLE$ & $\CIRCLE$ & $\CIRCLE$\\
        \citet{Fritz2014} 		& $\Circle$    & $\Circle$ & $\CIRCLE$ & $\LEFTcircle$ & $\Circle$\\
        \citet{Fritz2016} 		& $\Circle$    & $\Circle$ & $\CIRCLE$ & $\CIRCLE$ & $\Circle$\\
        \citet{Fucci2019} 		& $\Circle$    & $\Circle$ & $\CIRCLE$ & $\CIRCLE$ & $\Circle$\\
        \citet{Gonc2019}		& $\CIRCLE$    & $\Circle$ & $\CIRCLE$ & $\CIRCLE$ & $\Circle$\\
        \citet{Gui2019} 		& $\Circle$    & $\Circle$ & $\LEFTcircle$ & $\Circle$ & $\Circle$\\
        \citet{Radevski2015} 	& $\Circle$    & $\LEFTcircle$ & $\LEFTcircle$ & $\CIRCLE$ & $\Circle$\\
        \citet{Rostami2015} 	& $\Circle$    & $\LEFTcircle$ & $\LEFTcircle$ & $\CIRCLE$ & $\CIRCLE$\\ \bottomrule
    \end{tabular}
        
    \footnotesize
    \begin{tabular}{ccc}
        $\CIRCLE$ Completely Meets & $\LEFTcircle$ Partially Meets & $\Circle$ Don't Meets
    \end{tabular}
    
\end{table}

Therefore, the following research opportunity was identified: a holistic literature review that meets the five established comparison criteria. This opportunity is explored in the next sections.

\section{Methodology}
\label{sec:methodology}

This section aims to describe our review protocol. This protocol addresses essential steps defined in well established guidelines~\cite{Petersen2015} to plan and create a systematic mapping of the literature. In addition, our study protocol was also defined based on previously validated systematic mapping studies~\cite{bischoff2021technological,menzen2020using,carbonera2020software,gonccales2015comparison,gonccales2019comparison, bischoff2019integration, ROGER2020SBSI, souza2020big}



\subsection{Research Questions}

In addition to the objectives, the Research Questions (or RQs) will guide the researcher in the development of his systematic mapping, helping to identify possible studies that may converge with the theme of the study under development \cite{Petersen2015}. The Research Questions used in this study are presented in Table \ref{table:rq}, bringing their purposes and the variables associated with them.

Although systematic mapping has a well-established methodology for conducting it, its topics extend to the needs of the researcher \cite{Petersen2015}. Thus, the steps adopted for the development of the present work were structured as elucidated in subsequent sections.

\begin{table*}[ht]

    \caption{Research Questions, their description and related variables}
    \label{table:rq}
\small
    \centering
    \begin{tabular}{p{0.50\linewidth}p{0.30\linewidth}p{0.10\linewidth}}
        \toprule
            \textbf{Research Question} & \textbf{Description} & \textbf{Variable} \\
        \midrule
            \textbf{RQ01:} How would it be \textbf{taxonomy} to classify studies involving IDEs and psychophysiological data? & Aims to build up a taxonomy capable of classifying studies involving psychophysiological data and IDEs. & Taxonomy \\
            \textbf{RQ02:} What kind of \textbf{devices} has been used to collect psychophysiological data through studies? & List devices used in the selected studies to collect data over the experiments & Devices \\
            \textbf{RQ03:} What kind of \textbf{attributes} are extracted by devices used in the studies? & Elucidates the main objective of e the data types that have been extracted over the experiments and what it means within the scope of the study & Attributes\\
            \textbf{RQ04:} How the \textbf{integration} of extracted data are supported by the IDEs? & Explains how the extracted data can be integrated inside of an IDE to provide extraction of features. & Integration Support \\
            \textbf{RQ05:} For what \textbf{purposes} is the data being applied?? & Brings up what extracted data aims to explain in the IDEs' and software development contexts. & Purposes \\
            \textbf{RQ06:} What are the main contribution of the reviewed studies? & Finds out the contributions of each analyzed study in the context of psychophysiological data applied to software development process. & Contributions \\
            \textbf{RQ07}: What is the \textbf{nature} of the reviewed studies? & Understand what methodologies researchers are using within scope of applying  physiological data and its cognitive processes into IDEs and in the software development process & Nature \\
            \textbf{RQ08:} Which \textbf{publication venues} have the selected studies been released? & Identify venues where studies have been published across last ten years. & Publication Venues \\
        \bottomrule
        \end{tabular}
\end{table*}

\subsection{Search Strategy} 

The second step in the Systematic Mapping methodology is the definition of the search strategy, which aims to reach a representative sample of studies that can generically answer the Research Questions previously defined.

The Search Strategy used in this study was conceived in two stages: (1) selection of digital databases related to the study theme, and (2) construction of a search string and its variants, respecting the peculiarities of the search. search engines present in each database.

\subsubsection{Data Sources}\label{sec:db}

Table \ref{table:db} presents the electronic databases used. They were chosen because they are widely used and previous literature mapping studies have shown their effectiveness \cite{Bischoff2019, Goncales2019}. Also, database selection was based on the search engine coverage, which returned articles from leading Software Engineering journals.

\begin{table}[H]
	\caption{Databases and its digital addresses}
	\label{table:db}
	\footnotesize
	\centering
	\begin{tabular}{p{0.40\columnwidth}p{0.50\columnwidth}}
	\toprule
		\textbf{Database}            & \textbf{Address}         \\
	\midrule
		ACM Digital Library & http://portal.acm.org/       		\\
		DBLP                & https://dblp.uni-trier.de/        \\
		IEEE Xplore         & http://ieeexplore.ieee.org 		\\
		Science Direct      & https://www.sciencedirect.com/  	\\
		Scopus			    & https://www.scopus.com/     		\\
		Semantic Scholar     & https://www.semanticscholar.org/  \\
		Springer Link       & http://www.springerlink.com/   	\\
		\bottomrule
	\end{tabular}
\end{table}

\subsubsection{Search String}\label{sec:string}

The search string is a construct derived from the combination of major terms and their synonyms and is used as an input to database search engines. selected. This input allows search engines to return a collection of potentially relevant studies contributing to the development of Systematic Mapping. Table \ref{table:query} lists the main terms and their synonyms that were used in the development of this study.

To produce a functional search string, the following steps were taken to develop it: (1) define the main terms; (2) search for related synonyms or semantically equivalent words, and (3) match the main terms to their synonyms using logical operators such as `` AND '' and `` OR ''. Applying the characteristics mentioned, we arrived at the following search string:

\begin{table}[H]
\centering
    \begin{tabular}{|p{0.75\linewidth}|}
    \hline
           \textit{(EEG OR electroencephalography OR neural OR brain OR cognitive OR psychometric OR bioinformatic) AND (software OR developer OR programmer OR professional OR ``software engineering'') AND (ide OR code OR editor OR ``integrated development environment'')  }\\
       \hline
    \end{tabular}
\end{table}

Due to the syntactical peculiarities that come from each search engine contained in the databases, the initial search string needed to be individually tailored to provide a better result without losing its semantic value.

\begin{table}[ht]
	\caption{Search string major terms and their synonyms}
	\footnotesize
	\label{table:query}
    \centering
    \begin{tabular}{p{0.4\columnwidth}p{0.50\columnwidth}}
	\toprule
		\textbf{Main Term} 			& \textbf{Synonym} \\ 
    \midrule
		Neuroscience 			& eeg, fmri, electroencephalography, ``functional magnetic resonance imaging'', neural, brain, cognitive, psychometric, bioinformatic \\
		Software Engineering 	& software, developer, programmer, professional, software engineering \\
		IDE 					& ide, code, editor, integrated development environment \\
		Data Analysis 			& ``data analysis'', ``signal analysis'', analytic, ``signal processing'' \\
	\bottomrule
	\end{tabular}
\end{table}

\subsection{Inclusion and Exclusion Criteria}\label{sec:icec}

Once the databases were selected and the search string was constructed, the subsequent step in the Systematic Mapping development process was composed of the definition of the Inclusion Criteria and the Exclusion Criteria (CI and EC, respectively).

The Inclusion Criteria were applied to include studies in the filtering process after their initial collection in  databases~\cite{Petersen2015}. The Exclusion Criteria acted as a mechanism for discarding studies that did not represent answers. for Research Questions \cite{Petersen2015}. Therefore, the following items were defined as Inclusion Criteria:

\begin{itemize}[topsep=5pt]
    \item \textbf{IC01:} Published between the years of 2009 and 2019;
    \item \textbf{IC02:} Written in English;
    \item \textbf{IC03:} Available as full papers in digital databases;
    \item \textbf{IC04:} Related to the search string and research questions;
\end{itemize}

Furthermore, to filter previously selected studies by the Inclusion Criteria, the following Exclusion Criteria were applied:

\begin{itemize}[topsep=5pt]
    \item \textbf{EC1:} Match the keyword in the search string, but the context is different for research purposes;
    \item \textbf{EC2:} Papers with no abstract;
    \item \textbf{EC3:} Is just an abstract or only a summary, conference call;
    \item \textbf{EC4:} There is no relation with Software Engineering domain;
    \item \textbf{EC5:} Is a copy/duplicate or an older version of an already selected article;
    \item \textbf{EC6:} Was not possible to access the full paper;
    \item \textbf{EC7:} Motivation related;
\end{itemize}

\subsection{Data Extraction Strategy}\label{sec:extract}

Once the steps for selecting the studies were established, it was necessary to define two mechanisms, namely: (1) data extraction form and (2) classification schemes. Such mechanisms are of paramount importance for the development of works in the Systematic Mapping format, since, in a standardized way, they will define how the data contained in the selected works will be extracted and classified, thus enabling reliable identification of possible answers to the data previously defined in Research Questions \cite{Petersen2015}.

\subsection{Data Extraction Form}

\begin{figure*}
    \centering    
    \includegraphics[width=0.8\textwidth]{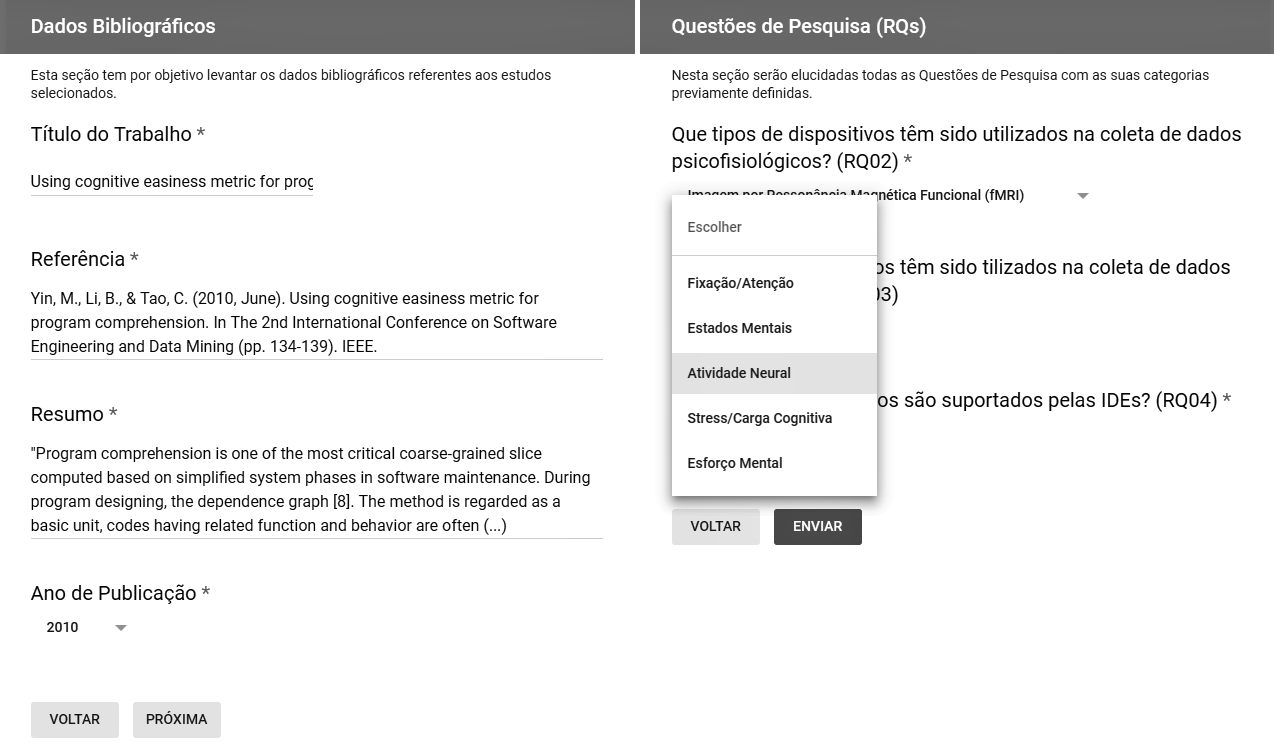}
    \label{fig:form}
    \caption{A snippet of data extraction form applied in the study development.}
\end{figure*}

To standardize the process of extracting the data contained in the articles and standardize the classification of the answers to the Research Questions, a data extraction form was developed, as can be observed in Figure \ref{fig:form}.

\subsubsection{Classification Schema}\label{sec:classification}

To develop an effective mechanism for classifying studies, among the possible approaches to category extraction used in the classification scheme, the following were adopted in this study:

\begin{itemize}
    \item \textbf{Topic Independent Scheme:} Generic categories are used to provide an overview of the selected studies and their theme. Usually, the authors recommend the use of categories already adopted in another systematic mapping;
    \item \textbf{Topic Dependent Scheme:} Specific questions emerge as the studies are filtered, thus bringing a detailed view of their content, which is a peculiarity present in each systematic mapping.
\end{itemize}

Based on the classification schemes presented, the following categories were extracted as possible answers for each Research Question: \newline

\begin{itemize}

 \item \textbf{Taxonomy (RQ01):} This category aims to guide the study in the construction of a taxonomy capable of assisting in the identification of studies that use psychophysiological data integrated with IDEs within the scope of Software Engineering. Unlike the other Research Questions, it was constructed as the answers of the others emerged.. \newline

\item \textbf{Devices (RQ02):} Represents which types of devices were used in the studies selected for the collection of psychophysiological data and other biomedical metrics. The following items represent the extracted categories used for the classification of studies in this regard:

\begin{itemize}
    \item Electroencephalogram (EEG): type of sensing used to measure and record brain electrical activity;
    \item Eye-tracker: a device used to measure the eye position of the individual as well as his movements;
    \item Functional Magnetic Resonance Imaging (fMRI): measures brain activity in association with changes in blood flow and variations in oxygen saturation;
    \item Functional Near-Infrared Spectroscopy (fNIRS): like fMRI, fNIRS collects brain activity measurements through hemodynamic responses;
    \item Electrodermal Activity (EDA): performs measurement series around galvanic skin variation;
    \item Device Combination: applicable when more than one device was used in the experiment, also applied to complementary sensing devices such as measuring bracelets and chest straps equipped with cardiac and respiratory monitoring sensors;
    \item No Devices: when the nature of the study did not require sensor devices to be used.
\end{itemize}

\item \textbf{Attributes (RQ03):} refers to context-based characteristics associated with data collected through sensor devices. The attributes identified in the selected studies were classified into:

\begin{itemize}
    \item Fixation/Attention: aims to indicate how long an individual's visual focus remains in an \textit{Area of Interest}, usually associated with attention \cite{Beelders2016}. However, without complementary Mental State measures, it is not possible to infer the relationship of this attribute to focused attention.
    
    \item Mental States: represents a set of attributes responsible for the relationship between cognitive processes (attention, learning, meditation, drowsiness, etc.) and the electrical activity of neurons, usually expressed through brain waves such as Alpha, Beta, Gamma, Delta and Theta \cite{Szu2013}.
    
    \item Neural Activity: this category covers sets of neural activation patterns and their associated semantics \cite{lane1997neural}.
    
    \item Stress/Cognitive Load: often associated with an individual's working memory as he or she performs a task or even a set of tasks that require context switching \cite{sweller1988cognitive}.
    
    \item Mental Effort: tt intends to explain the collection of biological metrics linked to the contribution of resources provided by the body during the execution of cognitive tasks \cite{Veltman1998a}.
    
    \item Combined Attributes: applicable to studies that have performed the analysis, three or more combined attributes.
    
    \item No Attributes: category adopted in studies in which specific attributes around psychophysiological data were not evaluated.
\end{itemize}

\item \textbf{Integration Support (RQ04):} this category indicates whether the study analyzed implemented any mechanism capable of integrating the collected data with the IDEs used.

\begin{itemize}
    \item Supported: applicable when the study analyzed used a tool capable of integrating the psychophysiological data collected and the IDE used;
    
    \item Not Supported: when a study used psychological data as well as IDEs, but data analysis was independent of IDE, needing an external tool to perform it;
    
    \item Not Applicable: category adopted in analyzed studies that use psychophysiological data but do not have in their context the use of IDEs;
\end{itemize}

\item \textbf{Purposes (RQ05):} it brings an overview of the main objectives stipulated by the authors in each study, regarding the use of the collected data and their related attributes. Thus, the identified categories were enumerated as follows:

\begin{itemize}
    \item Program Comprehension and Debugging Strategy: this category of purposes attempts to quantify the ease of understanding a program from the developer's point of view, in addition to analyzing the strategies adopted by it during the debugging process;
    
    \item Brain-Computer Interface: this category applied when the main purpose of the study is to develop an interface capable of enabling the interaction between humans and computers through the use of signals from neural activities;
    
    \item Cognitive Load: related to studies that conduct measurements linked to a developer's mental effort while performing a complex task;
    
    \item Improve Productivity: aims to find ways to provide better tools and processes in the context of Software Engineering and the quality of life of the developer;

    \item Out of Context: applicable where the purpose of the study was not directly linked to the use of psychophysiological data in the context of Software Engineering.
\end{itemize}

\item \textbf{Contributions (RQ06):} deviating from the others, this category adopts two approaches: the first one of subjective nature, will present a synthetic text with the main contributions identified by the authors for the development of this work. In the same vein, the second approach will seek to generalize the contributions identified through the following categories, according to the value returned for the development of the present study:

\begin{itemize}
    \item Metric: even indirectly, the study presents some metric derived from psychophysiological data that could be correlated to some of the identified attributes and purposes presented above;
    
    \item Tool: brought some tools for both data collection and processing/analysis;

    \item Model: when the study contributes to the presentation of a conceptual or practical model of architecture or experimentation involving the use of psychophysiological data;
    
    \item Method: presents an isolated method that can be used in the aspects of collecting, processing or analyzing psychophysiological data;

    \item Process: similar to the item \textit{method}, however, this category encompasses a sequence of methods for collecting, processing and or analyzing the collected data;

    \item Integration: It enables the identification of some approach of integrating the psychophysiological data collected with integrated development environments.

\end{itemize}

\item \textbf{Nature (RQ07):} the nature of a study refers to the type of research developed as to its main purpose and development. The classification style often adopted in the literature is the \cite{Wieringa2006}, which determine the following categories:

\begin{itemize}
    \item Proposal of Solution: relating to studies proposing new techniques for solving existing problems or revising techniques already adopted;
    
    \item Evaluation Research: this category applies to studies that evaluate techniques and practices applied in industry to specific problems;
    
    \item Validation Research: studies for the evaluation of innovative techniques, but not yet validated for their application in the industry;
    
    \item Philosophical Paper: studies which aim to bring a new perspective regarding the already consolidated knowledge structures;
    
    \item Personal Experience Paper: This type of work discussed with another author solved a problem by applying a technique in practice, without the need to also use the technique during the development of the study;
    
    \item Opinion Paper: it encompasses all studies that present opinions based on a particular research topic;
    
    \item Controlled Experiment: in this nature of studies, the authors performed controlled environment experiments following well-established and reproducible methodologies.

\end{itemize}

\item \textbf{Publication Venues (RQ08):} it identifies in which publications the selected studies have been published, such as conferences, workshops, scientific journals, etc., as the objective of inferring the maturity level of the topic in the scientific community.

\end{itemize}

\section{Threats to Validity}\label{sec:threats}

Regarding the development of Systematic Mappings, even if they have a well-defined and reproducible methodology, the threats to the validity of the study are inherent in the same \cite{Petersen2015}. In this regard, three types of threats have been enumerated: \textit{construction}, \textit{conduction} and \textit{conclusion}.

The first nature of threats -- \textit{construction} -- refers to the definitions present in the methodology, especially the search strings adapted for each database, which may not represent a return of the same value for each database. To mitigate this threat, individual search strings were built using the peculiarities of each search engine.

The subsequent nature of threats -- \textit{conduction} -- implies conducting the classification process, which may have adopted a very generic or, on the other hand, a very specific categorization schema, compromising the formation of study groups. In this aspect, an iterative approach was sought, that is, during the classification of articles, new categories were added to the exclusion of others, requiring new classification cycles of previously classified studies.

Finally, the \textit{conclusion} nature brings with it biases from the authors applied to the results, and may lead to convenient conclusions inclinations. Therefore, we sought to use the theoretical basis of similar studies to corroborate the results, thus reinforcing the conclusions obtained.

\section{Filtering Process Execution}\label{sec:filtering}

This section aims to describe how the filtering process of potentially relevant studies obtained from the electronic databases presented in Table \ref{table:db} of Section \ref {sec:db} was conducted.

\begin{figure*}[ht]
    \centering
    \includegraphics[scale=0.5]{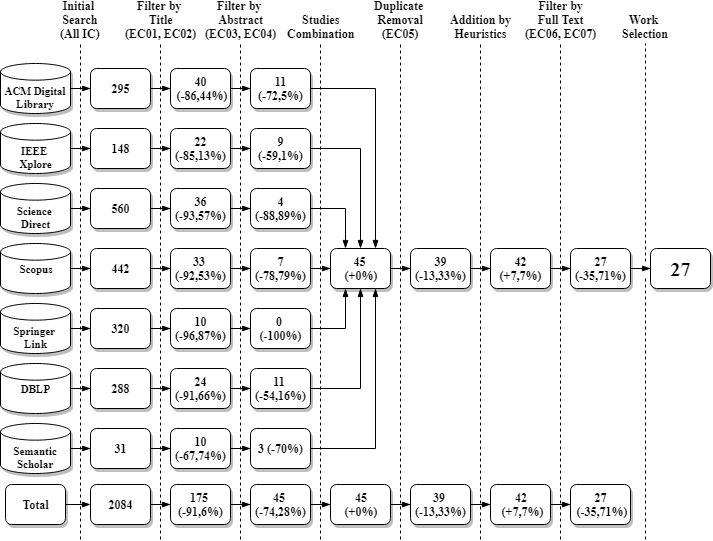}
    \caption{Study Filtering}
    \label{fig:filtering}
\end{figure*}

The filtering process used in this work took place in eight distinct steps, presented through Figure \ref{fig:filtering}. Except for the Initial Research stage, which only used the Inclusion Criteria in the screening, the others used groupings of the Exclusion Criteria as guidelines for the removal of the initially selected studies. The following are the steps taken as well as their driving guidelines.

\begin{itemize}

    \item \textbf{Initial Search:} this step was characterized by the insertion of the search string (Section \ref{sec:string}) into the chosen databases (Section \ref{sec:db}), combined with the composite filters parameterized according to the Inclusion (Section \ref{sec:icec}), with this, the search returned an amount of 2084 articles to be filtered, grouped according to their respective databases, as illustrated in Figure \ref{fig:filtering}.

    \item \textbf{Filter by Title:} at this stage of the filtering process, the titles of the articles resulting from the Initial Search were analyzed. Exclusion Criteria EC01 and EC02 were used for filtering (see Section \ref{sec:icec}), thus the initial amount of 2084 articles was reduced to a total of 175, representing a reduction of 91.6\% of the volume of studies initially selected (Figure \ref{fig:filtering}).
    
    \item \textbf{Filter by Abstract:} responsible for the accuracy in choosing relevant studies Systematic Mapping development, this step consisted of reading the abstracts of all 175 resulting articles. Exclusion criteria EC03 and EC04 (see Section \ref{sec:icec}) were applied as filters, reducing the volume of studies by 74.28\%, thus reaching a total of 45 articles (Figure \ref{fig:filtering}).
    
    \item \textbf{Studies Combination:} as illustrated in Figure \ref{fig:filtering}, the previous filtering steps applied to each of the electronic databases. Labels that linked articles to their original databases were removed, making them a single literary corpus, thus allowing the execution of the subsequent steps.

    \item \textbf{Duplicate Removal:} after article combination, the Exclusion Criteria EC05 was applied, responsible for removing duplicates of the studies and/or their old versions. Such filtering resulted in a reduction of 13.39\% of the total number of studies, thus resulting in a literary corpus of 39 articles (Figure \ref{fig:filtering}).
    
    \item \textbf{Addition by Convenience:} to increase the volume of studies analyzed, three articles related to the theme were added for convenience and within the Inclusion and Exclusion Criteria previously used, increasing by 7.7\% of the total or 42 articles (Figure \ref{fig:filtering}). Such addition was made by the authors' knowledge and through linked analysis of the references found in the articles already analyzed (snowballing), thus following the recommendations of \cite{Petersen2015} for selection.
    
    \item \textbf{Filter by Full Text:} in this step, the articles from the previous stage were read in full, applying the Exclusion Criteria EC06 and EC07 (Section \ref{sec:icec}) resulting in the removal of 35.71\% of the articles and generating an amount of 27 papers (Figure \ref{fig:filtering}).

    \item \textbf{Work Selection:} The final phase of the process resulted in the selection of 27 articles, named in this paper as Primary Studies, which are arranged in \ref{app:studies}. Besides, through the strategies described in Section \ref{sec:extract}, the data were extracted to obtain the answers to the Research Questions.
    
\end{itemize}

\section{Results}\label{sec:results}

This section presents the results obtained by executing the filtering process described in Section \ref{sec:filtering} according to the outline discussed in Section \ref{sec:methodology}. To present quantitative data for each Research Question, we used tables containing the Primary Study distribution, represented by their identifiers listed in \ref{app:studies}, according to their respective answer categories.

\subsection{RQ01: Taxonomy}

\begin{figure*}[ht]
    \centering
    \includegraphics[width=.8\linewidth]{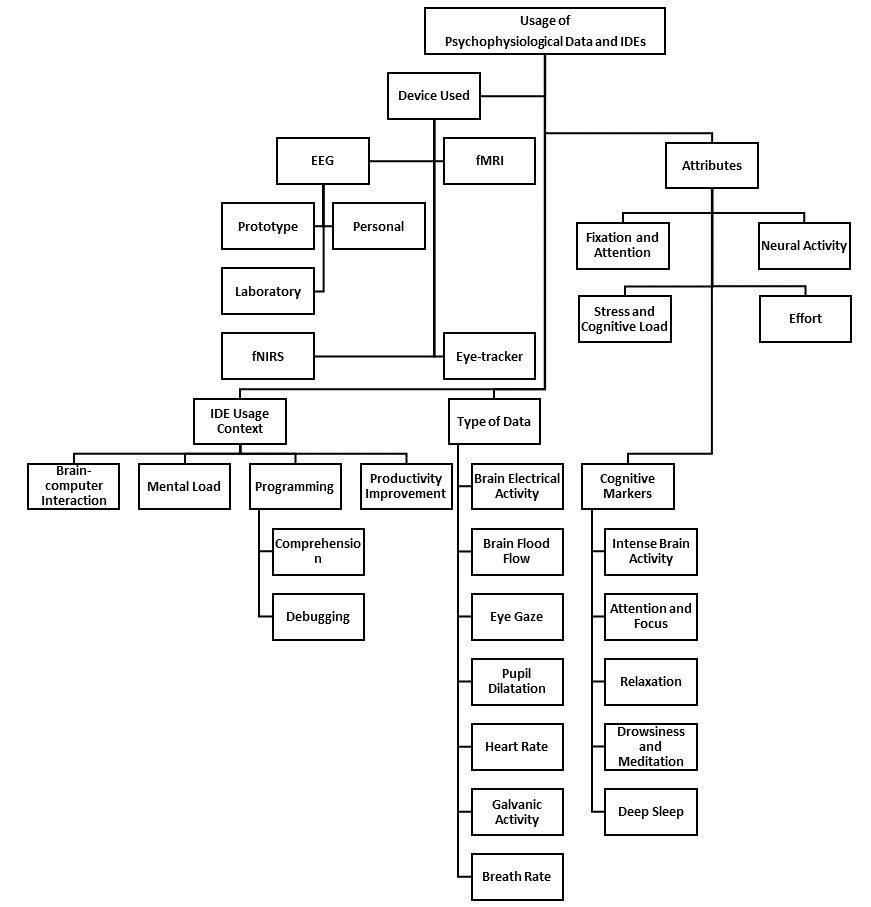}
    \caption {Taxonomy of psychophysiological data and IDEs usage.}
    \label{fig:taxo}
\end{figure*}

Figure \ref{fig:taxo} introduces a classification scheme (or taxonomy) produced after reading and examining the primary studies. The taxonomy structure has the variables explored in the formulated research questions, as well as the possible classifications (or values) for such variables found in the studies.

It is possible to observe that the second level of taxonomy was developed based on the classification criteria of the studies covered in the development of this work. The criteria are (1) devices that were used in the studies to collect psychophysiological data; (2) the context in which the data collected were used in the scope of the IDEs; (3) the types of data that were collected by the devices used and (4) what the collected data is capable of indicating.

Although taxonomy can bring an overview within the scope of this work and facilitate the process of classifying and identifying studies, it is worth noting that such a tool was developed to be applied in this study, therefore, it is necessary to validate it and/or adaptation when used in the development of similar works.

\subsection{RQ02: Devices}


Table \ref{tab:rq2} shows the results for RQ02. The main feature is that, although the most used device was the EEG (29.63\%, 8 of 27), there is no predominance of the devices used to capture psychophysiological data. Greater adoption of EEG could be explained by the improvement of data capture and mainly by the ease of use and cost reduction of such devices, thus helping in the wider use in a different context of studies~\cite{Rostami2015}. Furthermore, some studies~\cite {Rostami2015,Fritz2014,Fritz2016,Radevski2015} indicate that electroencephalography, through brain waves from neural activation, is suitable for identifying neural activities during the development of different tasks, including those related to software development, code understanding, and debugging strategies.

\begin{table}[!ht]
    \caption{List of devices used per study (RQ02)}
    \scriptsize
    \label{tab:rq2}
        \begin{tabular}{p{0.3\columnwidth} p{0.10\columnwidth} p{0.1\columnwidth} p{0.3\columnwidth}}
        \toprule
            \textbf{Device}                                        & \textbf{Studies} & \textbf{Percentage} & \textbf{Studies} \\ 
        \midrule
            N/A                                           & 3       & 11,11\%    & S15, S26, S27\\
            Eye-tracker                                   & 5       & 18,52\%    & S02, S03, S12, S16, S20 \\
            EEG                                           & 8       & 29,63\%    & S01, S04, S11, S14, S18, S19, S23, S25 \\
            fMRI                                          & 5       & 18,52\%    & S05, S08, S17, S21, S22 \\
            fNIRS                                         & 2       & 7,41\%     & S13, S24 \\
            Device Combination                            & 4       & 14,81\%    & S06, S07, S09, S10 \\
        \bottomrule
        \end{tabular}
\end{table}


Both \textit{eye-trackers} and \textit{fMRI} (Functional Magnetic Resonance Imaging) register 18.52\% of the results, followed by \textit{fNIRS} (Functional Near-Infrared Spectroscopy), which was found in 7.41\% of the studies.


The results also point out that some studies have chosen to use more than one type of device. The combination of devices was identified in 14.81\% of our sample. We observed that often the use of \textit{EEG} is combined with \textit{eye-trackers} or even complementary devices, such as measuring bracelets and chest bands equipped with cardio-respiratory frequency sensors. This combination seeks to complement the measurements in scenarios where \textit{eye-tracker} and \textit{EEG} cannot indicate, in isolation, attributes such as stress and focused attention \cite{Beelders2016}.


Given the nature of some of the Primary Studies, as well as the methodology adopted, the use of devices for data collection was not found. Such a range of studies represented 11.11\% of our sample.

\subsection{RQ03: Attributes}

Regarding the analyzed attributes, Table \ref{tab:rq3} clarifies their distributions in Primary Studies. There is a predominance (29.63\%) of studies that had the subject's neural activity as an object of study. Such works used devices such as EEGs, fNIRS, and fMRI to evaluate which brain regions were activated during the execution of their experiments, in addition to relating certain semantics to the observed brain activities \cite{lane1997neural}.

\begin{table}[!ht]
    \caption{Distribution of attributes analyzed in each selected study (RQ03)}
    \label{tab:rq3}
    \scriptsize
        \begin{tabular}{p{0.32\linewidth} p{0.10\linewidth} p{0.1\linewidth} p{0.25\linewidth}}
        \toprule
            \textbf{Attribute} & \textbf{Studies} & \textbf{Percentage} & \textbf{Studies}\\ 
        \midrule
            N/A						& 3       & 11,11\%    & S15, S26, S27\\
            Fixation/Attention		& 5       & 18,52\%    & S02, S03, S12, S16, S20\\
            Mental States			& 7       & 25,93\%    & S01, S04, S14, S18, S19, S23, S25\\
            Neural Activity			& 8       & 29,63\%    & S05, S08, S11, S13, S17, S21, S22, S24\\
			Attributes Combination 	& 2       & 7,40\%     & S09, S10\\
            Fixation/Attention, Neural Activity & 2       & 7,40\%     & S06, S07\\ 
        \bottomrule
        \end{tabular}
\end{table}

Studies that have lectured around \emph{Mental States} accounted for 25.93\% of Primary Studies. In these studies, we opted primarily for the use of EEGs for the observation of mental states, given their intimate connection to brain waves (Alpha, Beta, Gamma, Delta, and Theta), in addition to being a topic that was very present in the specialized literature \cite{Szu2013}.

The \emph{Fixation / Attention} was the attribute studied in 18.52\% of the studies present in the literary \textit{corpus} adopted in this systematic mapping. Usually, the fixation was attributed by the studies to the ocular focus at points of interest, predominantly measured by \emph{Eye-trackers}, see data presented in Table \ref{tab:rq2} however, in cases where attention was the object studies, complementary sensors were used concurrently.

In 7.40\% of the works, a combined analysis of the attributes of \emph{Fixation / Attention} and \emph{Neural Activity} was performed. In these studies, it was observed that the authors' focus permeated the understanding of source code, for that, \emph {Eye-trackers} were used to collect data related to ocular positioning and / or focus and fNIRS, to identify neural activation of brain areas linked to learning and reading \cite{Fakhoury2018, Fakhoury2018a}.

Studies that analyzed more than three attributes in their experiments correspond to 7.40\% of the total. In these works, situations of greater specificity were used when compared to others, such as the measurement of \emph{Fixation / Attention}, \emph{Mental States} and complementary attributes (cardiorespiratory frequency) during the execution of daily tasks by a developer \cite{Fritz2014}, or even the combination of \emph{Mental States} with attributes from myoelectric signals (blink of the eye, temporomandibular tension), to use them to compare the level of understanding of source code between developers new and experienced \cite{Lee2016}.

Although 88.88\% of the studies applied analyzes of attributes related to the collected psychophysiological data, in 11.12\% of them there was no analysis of these attributes. This characteristic was due to the nature of these works, in addition to the methodological process adopted in their development.

\subsection{RQ04: Integration Support}

\begin{figure}[H]
    \centering
    \includegraphics[width=0.5\textwidth]{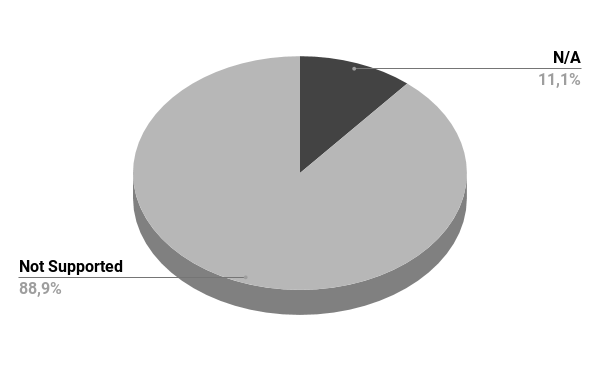}
    \caption{Data integration supporting by IDEs across studies (RQ04)}
    \label{fig:rq4}
\end{figure}

This Research Question aimed to obtain as an answer what types of support the IDEs used in the experiments provided for the integration of the collected psychophysiological data, thus enabling its analysis in the context of Software Engineering by the researchers. Figure \ref{fig:rq4} illustrates how the answers for this Research Question were divided.

It is observed that, of the total of 27 studies analyzed, 11.1\% of these made a type of use of psychophysiological data, however, without the use of IDEs in their experiments, as previously described in Section \ref{sec:classification} and illustrated through Figure \ref{fig:rq4}. On the other hand, about 88.9\% of the studies were classified in the \emph{Unsupported} category, that is, they carried out experiments involving IDEs and also the analysis of psychophysiological data. Still, none of these studies specified mechanisms capable of integrating the collected data and the Integrated Development Environments (IDEs), therefore, the analyzes were carried out externally and using specialized tools.

\subsection{RQ05: Purposes}

In this item, the purposes adopted by the authors in the Primary Studies regarding the use of psychophysiological data in the context of Software Engineering are set out. Through the data elucidated in Table \ref{tab:rq5}, it was found that, in the majority (66.70\%), the Primary Studies focused on analyzing the \emph{Program Comprehension and Debugging Strategy} used by the developers in the experiments performed.

\begin{table*}[!ht]
\caption{Purpose of data analysis in each study (RQ05)}
\label{tab:rq5}
\scriptsize
        \begin{tabular}{p{0.5\linewidth} p{0.10\linewidth} p{0.1\linewidth} p{0.2\linewidth}}
\toprule
\textbf{Purpose}                                                      & \textbf{Studies} & \textbf{Percentage} & \textbf{Studies} \\ \midrule
Program Comprehension and Debugging Strategy 		& 18        & 66,70\%    & S02, S03, S05, S06, S07, S08, S09, S12, S13, S14, S16, S17, S20, S21, S22, S25, S26, S27\\
Mental Workload                                     & 4         & 14,81\%    & S04, S11, S19, S24\\
Human-Computer Interaction                          & 2         & 7,41\%     & S01, S23\\
Improvement in Productivity                         & 2         & 7,41\%     & S10, S18\\
N/A				                                    & 1         & 3,70\%     & S15 \\
\bottomrule
\end{tabular}
\end{table*}

It was also observed that, in 14.81\% of the Primary Studies, the focal purpose of the work was the evaluation of the developer while he performed a series of activities related to his context. It is worth noting that, despite the possibility of evaluating Source Code Understanding and the debugging strategy individually, the measurements related to these purposes have greater validity when linked to the \emph{Cognitive Load}, according to surveys developed by others authors \cite{Gonc2019}.

Constituting 7.41\% of the Primary Studies were those whose main purpose was the search for the \emph{Improvement in Productivity} of the developers through the use of psychophysiological data and IDEs.

Studies involving the development of \emph{Brain Computer Interfaces} also corresponded to 7.41\% of Primary Studies. This category includes the construction of a device prototype for the control of home peripherals utilizing brain waves \cite{Szu2013} and the idealization of a portable EEG capable of communicating with \textit {smartphones} \cite{Alshbatat2014}.

\subsection{RQ06: Contributions}\label{sec:result_rq6}

This section aims to present the main contributions identified in each of the Primary Studies. Such contributions were extracted during the reading of the full text (Section \ref{sec:filtering}) and aimed to answer this Research Question. The texts were synthesized and arranged through the following list, to allow a structured reading, even if the source of the data was qualitative and with subjective aspects.

\begin{itemize}
	
	\item \textbf{S01} and \textbf{S18}: Enabled the acquisition and processing of neural activity signals in real-time, in addition to the extraction of features to use such signals as BCI (\textit {Brain-computer Interface}).
	
	\item \textbf{S02}: It brought a vision of how the display of code information is interpreted by both novice and experienced programmers.
	
	\item \textbf {S03}: The study indicated that, although the fixation time is longer in cases where the code is presented in black and white, compared to the same section with syntax highlighting, this difference does not is significant in terms of understanding the code.
	
	\item \textbf{S04}: In this study, it is worth highlighting two main contributions: (1) the methodology for using EEG in the assessment of cognitive load and (2) the evidence that more experienced programmers solve tasks more quickly, generating a lower cognitive load in the same amount of time as beginners.
	
	\item \textbf {S05}: As the main contribution of this study, the authors have brought a mapping of brain activation during the clearance process.
	
	\item \textbf {S06}: The authors proposed methodologies to relate the developer's cognitive load to the low quality of the source code.
	
	\item \textbf {S07}: This study was important for bringing a review of how other authors are using the Eye-tracker and fNIRS as tools for measuring the cognitive/mental load of developers in terms of understanding the source code.
	
	\item \textbf{S08}: Among the contributions of this study, two are of great value for the development of this mapping: (1) it clarifies that the programming language and the natural language are interpreted in different regions of the brain and (2) with the increase in the programmer's expertise, the programming language begins to be interpreted as a natural language by your brain.
	
	\item \textbf{S09}: The work brought as a contribution to the possibility of using neurophysiological signals to assess the difficulty in programming activities, in addition to bringing the possibility of developing tools to improve the software development process.
	
	\item \textbf{S10}: The authors brought a revision of great value around psychophysiological data and their sensing devices. They also presented the techniques used to analyze the data used and the algorithms for predicting similar cases.
	
	\item \textbf{S11}: As a result, the authors have shown that with increasing cognitive load, productivity degrades over time.
	
	\item \textbf{S12}: Although an experiment was carried out in the context of the use of programming languages, the study presented the participants' eye movements as a result, without drawing parallels with the literature to find patterns or explanations for the identified patterns.
	
	\item \textbf{S13} and \textbf{S14}: The results are useful in understanding what happens in the programmer's brain when observing certain aspects (logic, code flow) during the process of understanding programming languages. By comprehension of the brain's activation patterns, it is possible to create metrics for assessing the skills of both novice and experienced programmers. It is still possible to determine the factors that make teaching programming more effective in terms of increasing cognitive ability.
	
	\item \textbf{S15}: A working context-sensitive IDE has been developed and tested by the authors.
	
	\item \textbf{S16} and \textbf{S20}: The authors exposed the state-of-the-art around the use of eye-tracking in Software Engineering, in addition to advances in technology over the years.
	
	\item \textbf{S17}: Developed guidelines for establishing fMRI as a standard tool for observing the understanding of source code and other related processes.
	
	\item \textbf{S21}: As a contribution of the study, the authors showed that there is an activation pattern associated with working memory and natural language processing. As a contribution, the work brought a methodology for using fMRI in the context of understanding programming languages.
	
	\item \textbf{S22}: The authors identified that the use of semantic suggestions reduces brain activation during the process of understanding the source code, in addition to reproducing the studies of other authors and validating their results.
	
	\item \textbf{S23}: The work in question brought the feasibility of developing a domestic EEG, both for use as monitoring equipment and for BCI, given the possibility of analyzing data in real-time. It also elucidated the mental states linked to the brain wave spectrum.
	
	\item \textbf{S24}: The present study contributed with the evidence that, during the teaching process of object-oriented programming, visual IDEs such as BlueJ, reduce the cognitive load, making it possible to infer a more facilitated learning process. The authors also encourage studies on the use of IDEs and fNIRS.
	
	\item \textbf{S25}: Although the work evaluated the behavior of the Theta and Alpha waves, the authors did not bring a psychophysiological perspective of the cognitive processes involved during the activities.
	
	\item \textbf{S26}: He proposed a metric to assess cognitive complexity based on a graph of method calls.
	
	\item \textbf{S27}: The authors showed that the debugging methodologies used by developers vary according to the features of the IDEs, as opposed to traditional forms of debugging.
\end{itemize}

In addition to the results presented in the form of synthetic texts, the metrics were arranged according to the categories to which the respective contributions were classified, as shown in Table \ref{tab:rq6}.

\begin{table}[ht]
	\small
	\centering
	\caption{Distribution of studies by type of Contribution (RQ06)}
	\scriptsize
	\label{tab:rq6}
        \begin{tabular}{p{0.2\columnwidth} p{0.10\columnwidth} p{0.1\columnwidth} p{0.4\columnwidth}}

\toprule
\textbf{Purpose} & \textbf{Studies} & \textbf{Percentage} & \textbf{Studies} \\ \midrule
		Metric	    	& 10    & 37,04\%   & S03, S04, S05, S10, S12, S13, S14, S21, S22, S26\\ 
		Tool			& 2     & 7,41\%	& S09, S15\\ 
		Model	    	& 6     & 22,22\%	& S01, S16, S18, S19, S20, S23\\
		Method	    	& 6     & 22,22\%	& S02, S06, S07, S08, S24, S25\\ 
		Process			& 3	    & 11,11\%	& S11, S17, S27\\ 
		\bottomrule
	\end{tabular}
\end{table}

There is a greater number (37.04\%) of studies that contributed significantly by providing some kind of \emph{Metrics} about the use of psychophysiological data applied to the attributes listed in RQ03 and purposes of RQ05. Therefore, it is observed that \emph{Method} and \emph{Model} presented the same frequencies (22.22\% each), bringing both contributions that could be applied to the definition of an architectural model or experimental method. 

The category \emph{Process} brought contributions (11.11\%) to the definition of processes that could be adopted concerning the collection, processing, and analysis of psychophysiological data. Finally, only 2 studies (7.41\%) presented any tool that could be applied in the context of this work.

It is worth noting that none of the studies analyzed were classified in the category \textit{Integration} since they did not present any mechanisms capable of supporting the integration of the data collected with the integrated development environments or other tools used.

\subsection{RQ07: Nature}

In this Research Question, we sought to provide an overview of the distribution of each selected study, using the criteria defined in Section \ref{sec:classification}. The results obtained by the process of conducting systematic mapping are organized and arranged in Table \ref{tab:rq7}.

\begin{table}[ht]
	\scriptsize
	\centering
	\caption{Distribution of studies by Nature (QP07)}
	\label{tab:rq7}
        \begin{tabular}{p{0.2\columnwidth} p{0.10\columnwidth} p{0.1\columnwidth} p{0.4\columnwidth}}

		\toprule
		\textbf{Nature} & \textbf{Studies} & \textbf{Percentage} & \textbf{Studies} \\ 
		\midrule

		Controlled Experiment	& 15    & 55,56\%  & S02, S03, S04, S05, S06, S08, S09, S10, S11, S12, S13, S14, S21, S22, S25  \\ 
		Proposal of Solution    & 7     & 25,93\%  & S07, S15, S18, S23, S24, S26  \\
		Evaluation Research     & 1     & 3,70\%   & S19  \\
		Opinion Article		    & 4     & 14,81\%  & S16, S17, S20, S27  \\
		\bottomrule
	\end{tabular}
\end{table}

Note that predominantly the selected studies (55.56\%) are of nature \emph{Controlled Experiment}, that is, studies in which the authors perform some type of formatting experiment to validate an existing technique or confirm a concept found in the literature. Next are the studies of nature \emph{Proposal of Solution}, representing 25.93\% of the total value. In this nature of the studies, are listed those that the main objective was to present a solution to an existing problem or an improved version of a solution already published, through a proof of concept. The \emph{Opinion Article} articles 14.81\%, brings works that express an opinion of the authors on specific subjects, instigating a discussion and opening an opportunity for new works. Finally, 3.70\% of the works were of a nature \emph{Evaluation Research}, where you can evaluate a tool, methodology, or idea.

\subsection{RQ08: Publication Venues}

This Research Question aimed to quantify the venues in which the Primary Studies were published. The data regarding the responses acquired through the development of this study are shown in Table \ref{tab:rq8}

\begin{table}[H]
	\caption{Distribution of venues where studies were published (RQ08)}
	\label{tab:rq8}
	\scriptsize
        \begin{tabular}{p{0.2\columnwidth} p{0.10\columnwidth} p{0.1\columnwidth} p{0.4\columnwidth}}
		
		\toprule
		\textbf{Venue}      & \textbf{Studies} & \textbf{Percentage} & \textbf{Studies} \\ 		
		\midrule
		
		Conference & 16    & 59,26\%    & S03, S04, S06, S07, S08, S09, S10, S11, S12, S13, S14, S15, S17, S21, S25, S26\\
		Journal    & 6     & 22,22\%    & S01, S02, S16, S20, S23, S24\\
		Workshop   & 2     & 7,41\%     & S18, S19\\
		Symposium  & 3     & 11,11\%    & S05, S22, S27\\
		\bottomrule
	\end{tabular}
\end{table}

It appears that in its majority (59.26\%), the Primary Studies were published in \emph{Conferences} related to the main area, indicating a greater incipience about the theme of the selected works. Consequently, there are the works that were published in \emph{Journals} (22.22\%), featuring subjects consolidated in the scientific community. Finally, works that have been published in smaller events are presented, such as \emph{Symposium} (11.11\%) and \emph{Workshops} (7.41\%).

\subsubsection{Publication Frequency}

\begin{figure}[H]
    \centering
    \includegraphics[width=0.5\textwidth]{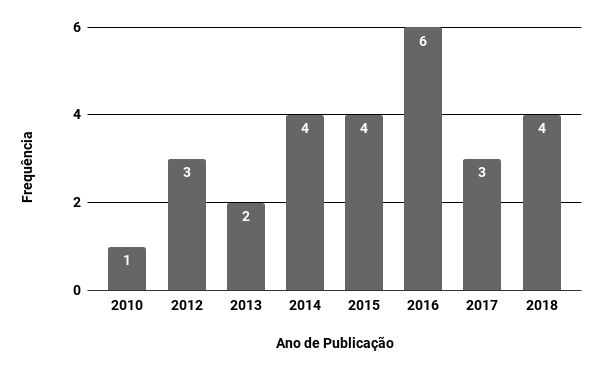}
    \caption {Publication of selected studies over the years }
    \label{fig:years}
\end{figure}

In addition to the results presented around the \emph {Publication Venues}, a survey regarding the distribution of the published studies over the years was also carried out, as quantified through Figure \ref{fig:years}.

It is possible to verify that, during the last decade -- period evaluated in the development of this systematic mapping -- there was an increase in the number of publications regarding the use of psychophysiological data applied in the context of Software Engineering.

\section{Discussion and Challenges}\label{sec:discussion}

\begin{figure*}[t]
	\centering
	\caption{Density of studies due to their nature, purpose and contributions.}
	\includegraphics[scale=0.4]{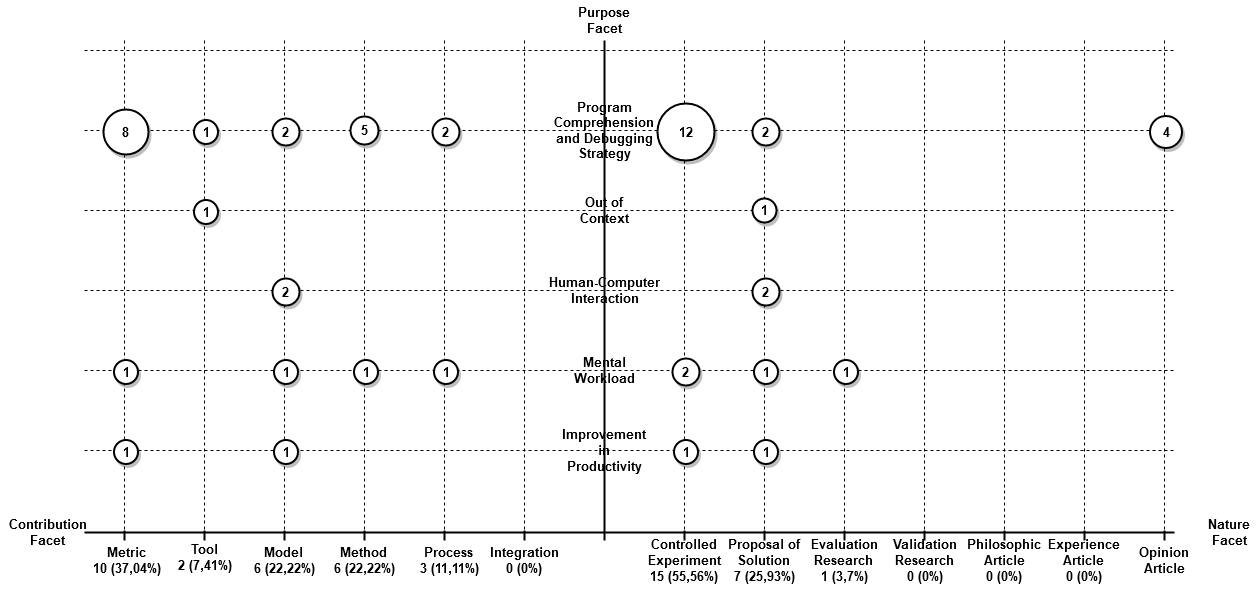}\\
	\label{fig:bubble_rq6}
\end{figure*}

\begin{figure*}[t]
	\centering
	\caption{Heatmap showing study density for each pair of responses to the applied Research Questions.}
	\includegraphics[scale=0.52]{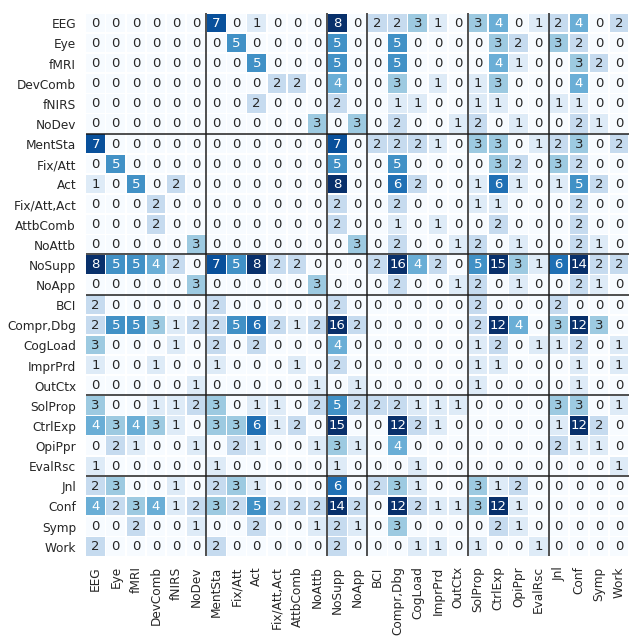}\\
			\footnotesize \underline{Legend:} \\ \textbf{EEG:} Electroencephalography -- \textbf{Eye:} Eye-tracker -- \textbf{fMRI:} functional Magnetic Resonance Imaging -- \textbf{DevComb:} Device Combination -- \textbf{fNIRS:} functional Near-Infrared Spectroscopy -- \textbf{NoDev:} No Device -- \textbf{MentSta:} Mental States -- \textbf{Fix/Att:} Fixation/Attention -- \textbf{Act:} Neural Activity -- \textbf{Fix/Att,Act:} Fixation/Attention, Neural Activity -- \textbf{AttComb:} Attributes Combination -- \textbf{NoAtt:} No Attributes -- \textbf{NoSupp:} Not Supported -- \textbf{NoApp:} Not Applicable -- \textbf{BCI:} Brain-computer Interaction -- \textbf{Compr,Dbg:} Program Comprehension and Debugging Strategy -- \textbf{CogLoad:} Cognitive Load -- \textbf{ImprPrd:} Productivity Improvement -- \textbf{OutCtx:} Out of Context -- \textbf{SolProp:} Proposal of Solution -- \textbf{CtrlExp:} Controlled Experiment -- \textbf{OpiPpr:} Opinion Article -- \textbf{EvalRsc:} Evaluation Research -- \textbf{Jnl:} Journal -- \textbf{Conf:} Conference -- \textbf{Symp:} Symposium -- \textbf{Work:} Workshop
	\label{fig:heatmap}
\end{figure*}

Through Figure \ref{fig:bubble_rq6} a bubble chart is presented where the primary studies are organized into three facets: contribution (left horizontal axis), purposes (vertical axis), and nature (right horizontal axis). Such a chart brings a combined view of the primary studies to identify the relationships between the contributions of each study, their nature, and the purposes adopted. The heat map illustrated in Figure \ref {fig:heatmap}, on the other hand, elucidates the density of studies for each pair of answers to the research questions presented in this work. Therefore, using these resources, it was possible to identify the aspects discussed in the following paragraphs.

\emph{Lack of Support for Integrating Psychophysiological Data:} Studies that carried out experiments using integrated development environments and psychophysiological data showed that there are no native mechanisms to support such types of data on the part of IDEs, nor through extensions that add such functionality. In this respect, data were collected by proprietary tools that communicated with the sensor devices, to be subsequently pre-processed and analyzed. Such a lack of support for integration proves to be opportune for the use of \emph{Big Data} tools that work with large volumes of data and/or processing of data flows in real-time. Furthermore, the production of infrastructure technology for the integration of psychophysiological data into integrated development environments through extensible functionalities is shown to be a necessity present in experimental studies applied to Software Engineering.

\emph{Source Code Understanding Experiments and Debugging Strategies Applied to Software Engineering:} Most of the authors of the studies reviewed in this work adopted the main purpose of using psychophysiological data to assess source code understanding. by the developers, in addition to their strategies adopted for the debugging process. This preference is elucidated by the domain of the studies analyzed, which usually sought applicability in the context of Software Engineering. Analyzing Figure \ref{fig:heatmap}, it becomes evident that this type of purpose was adopted in studies that used controlled experiments as the main methodology.

\emph{Data Analysis Using External Tools:} Although studies have performed data analysis to obtain the results and develop their discussions and conclusions, there was a need for analysis external to the controlled environment of the experiment, through the use of tools such as MATLAB, or proprietary solutions developed by the suppliers of the sensor devices used in the experiments. This characteristic was evident in the experiments in which integrated development environments were used (see item \textit {NoSupp} of Figure \ref{fig:heatmap}). In these experiments, in addition to the data collected by the devices, it was demanded that the metrics obtained in the IDEs were independently correlated to the psychophysiological data. In the aspect presented, there is an opportunity for the development of extensions to IDEs that enable the use of data analytics techniques in a native way, thus allowing the extraction of insights around psychophysiological data integrated to IDEs.

\emph{Increasing of Interest in Psychophysiological Data Applied to Software Engineering:} It is possible to observe a greater presence of studies published in conferences related to applied computing. This characteristic suggests the incipience of this research topic, given the interest in the development of works of this kind and the increase in publications in recent years (Figure \ref{fig:years} in Section \ref{sec:results}). Notwithstanding, the multidisciplinary integration between Neuroscience and Software Engineering opens scope for the production of diverse and innovative studies regarding the applicability of psychophysiology combined with the software development process. This phenomenon becomes observable through the plurality of contributions (Item \ref{sec:result_rq6}) from the analyzed articles, as well as by the related works used (Item \ref{sec:related_works}), which served as a theoretical basis during the production of this systematic mapping.

\section{Conclusions and Future Works}\label{sec:conclusion}

The development of this work provided the authors with a theoretical glimpse of the current state of the art regarding the use of psychophysiological data applied to Software Engineering, since its acquisition, processing, and extraction of insights on the physical and psychological processes involved in the production of software artifacts by the developer. It was possible to observe that, in the majority, the authors opted for the use of electroencephalography (29.63\%), eye-tracking (18.52\%) and functional magnetic resonance imaging (18.52\% ) in their experiments, this fact is corroborated by the maturity of the use of equipment present in the literature, as well as the reduction of costs and ease in the configuration, operation, and extraction of data from the measurements.

An expressive number of studies were found that analyzed the principles of cognitive load theory which, as well as the use of electroencephalography and eye-tracking, demonstrated to be well consolidated in the specialized literature, reinforcing the theoretical basis of the authors, especially in experiments that aimed to understand how the developer understands the source code and what strategies it uses during the debugging process (66.7\%), as well as in those that sought to assess the levels of the cognitive load presented by developers during the execution of their tasks (14.81\%).

A point that could be observed when reading the articles in the filtering and selection stage, was the lack of consensus of the authors regarding possible cognitive processes related to the developer's activities, except for the cognitive load, other concepts when applied in the context of Software Engineering, they presented an ambiguous content in relation, as is the example of neural activity, in which some studies conceptualized it as being the intensity of electrical and/or hemodynamic activation in the brain, while other studies presented this concept relating only with punctual areas of the brain and assigning semantics to it.

In the aspect of data analysis, the authors mainly used external tools for processing, extracting features, and obtaining insights about the data sets, especially in studies of an experimental nature in which the authors applied the manipulation of integrated development environments as part of the data sources. In these studies, no ways of integrating data from IDEs with analysis tools were presented, requiring the use of generic third-party software suites for the manipulation, processing, and analysis, or tools provided by the developers of the sensor devices used (88.9\%). This factor proved to be an opportunity for the application of multidisciplinary knowledge, such as the use of Big Data tools to manipulate the volume of data and, real-time processing or near real-time, extracting insights using machine learning algorithms, making wearable and Internet of Things devices to expand data collection outside a specific environment, enabling understand the psychophysiological characteristics presented by the developer in his real work environment, among other applications.

Through the opportunities observed during the production of this systematic mapping, it is intended to develop, as future works, mechanisms that enable the collection and processing of psychophysiological data from sensor devices, as well as their integration into integrated development environments, in a way to enable the generation of metadata that can be related to the events generated within the IDEs. In possession of such data, it will still be possible to apply machine learning techniques to obtain patterns that may suggest potential failures in the production of software according to the psychophysiological standard of each developer, enabling the creation of intelligent and adaptable IDEs to the user, in addition to tactical resources for prioritizing refactoring in software units.


\section{Acknowledgments}

This work was supported by the Coordena\c{c}\~{a}o de Aperfei\c{c}oamento de Pessoal de N\'{i}vel Superior – Brasil (CAPES) under Grant 001; National Council for Scientific and Technological Development (CNPq) under Grant 313285/2018-7.

\bibliography{bibliography}
\bibliographystyle{elsarticle-harv}

\appendix

\section{Search String per Database} \label{app:strings}
\begin{table}[ht]
	\footnotesize
	\centering
	\begin{tabular}{|C{0.15\linewidth}|J{0.75\linewidth}|}
		\hline
		\multicolumn{1}{|c}{\textbf{Database}} & 
		\multicolumn{1}{|c|}{\textbf{Search String}}\\
		\hline
		ACM Digital Library         & acmdlTitle: (    
		+(eeg fmri electroencephalography neural brain cognitive psychometric  bioinformatic ``functional magnetic resonance imaging'')     
		+(software developer programmer professional)     +(ide code editor ``integrated development environment'' ``software engineering'')) OR recordAbstract: (    +(eeg fmri electroencephalography neural brain cognitive psychometric  bioinformatic ``functional magnetic resonance imagin'')     +(software developer programmer professional)     +(ide code editor ``integrated development environment'' ``software engineering'')) \\ \hline
		IEEE Xplore                 & (sofware OR developer OR programmer OR professional OR ``software                             engineering'') AND (eeg OR fmri OR electroencephalography OR neural                             OR brain OR cognitive OR psychometric  OR bioinformatic OR                                     ``functional magnetic resonance imaging'') AND (ide OR code OR                                  editor OR ``integrated development environment'')\\ \hline
		Science Direct              & (\{``software engineering''\} OR developer OR programmer OR code)                              AND (IDE ) AND (eeg OR Fmri OR cognitive OR bci)\\ \hline
		Scopus                      & ALL ( ( ``software engineering''  OR  developer  OR  programmer  OR                             code )  AND  ( ide  OR  ``integrated development environment'' )  AND                             ( eeg  OR  fmri  OR  cognitive  OR  bci ) )  AND  PUBYEAR                                      \textgreater  2008  AND  ( LIMIT-TO ( SUBJAREA ,  ``COMP'' )  OR                                 LIMIT-TO ( SUBJAREA ,  ``ENGI'' ) )  AND  ( LIMIT-TO ( LANGUAGE ,                                ``English'' ) )\\ \hline
		Springer Link               & (``integrated development environment'' OR IDE) AND (``software                                 engineering'' OR ``software development'' OR programmer OR developer)                             AND (bci OR eeg OR fmri OR cognitive)\\ \hline
		DBLP                        & IDE|``integrated development environment''\\ \hline
		Semantic Scholar            & (IDE OR ``integrated development environment'') \\ \hline 
	\end{tabular}
\end{table}

\section{Selected Studies} \label{app:studies}
\begin{enumerate}
	\item [S01] \citet{Alshbatat2014}
	\item [S02] \citet{Bednarik2012} 
	\item [S03] \citet{Beelders2016}
	\item [S04] \citet{Crk2014}
	\item [S05] \citet{Duraes2016}
	\item [S06] \citet{Fakhoury2018}
	\item [S07] \citet{Fakhoury2018a}
	\item [S08] \citet{Floyd2017}
	\item [S09] \citet{Fritz2014}
	\item [S10] \citet{Fritz2016}
	\item [S11] \citet{Funke2013}
	\item [S12] \citet{Hejmady2012}
	\item [S13] \citet{Ikutani2014}
	\item [S14] \citet{Lee2016}
	\item [S15] \citet{Logozzo}
	\item [S16] \citet{Obaidellah2018}
	\item [S17] \citet{Peitek2018}
	\item [S18] \citet{Radevski2015}
	\item [S19] \citet{Rostami2015}
	\item [S20] \citet{Sharafi2015a}
	\item [S21] \citet{Siegmund2014}
	\item [S22] \citet{Siegmund2017}
	\item [S23] \citet{Szu2013}
	\item [S24] \citet{Uysal2016}
	\item [S25] \citet{Yeh2017}
	\item [S26] \citet{Yin2010}
	\item [S27] \citet{Zayour2016}
\end{enumerate}

%
%

%
%

\end{document}